\documentclass[10pt,conference]{IEEEtran}
\IEEEoverridecommandlockouts
\usepackage{amsmath,amssymb,amsfonts}
\usepackage{algorithmic}
\usepackage{graphicx}
\usepackage{textcomp}
\usepackage{xcolor}
\usepackage{xspace}
\def\BibTeX{{\rm B\kern-.05em{\sc i\kern-.025em b}\kern-.08em
    T\kern-.1667em\lower.7ex\hbox{E}\kern-.125emX}}
    
\usepackage[english]{babel}
\usepackage[numbers]{natbib}
\usepackage{balance}
\usepackage[hidelinks]{hyperref}
\usepackage[inline, shortlabels]{enumitem}
\usepackage{mathrsfs}
\usepackage[edges]{forest}

\usepackage{booktabs}% http://ctan.org/pkg/booktabs
\usepackage{tabularx}
\usepackage{adjustbox}
\usepackage[figuresright]{rotating}
\usepackage{multirow}
\usepackage{wrapfig}
\usepackage[most]{tcolorbox}

\usepackage{amssymb}
\usepackage{amsthm}
\usepackage{amsmath}%,amsfontshttps://www.overleaf.com/project/62558251243dd466c5c72bcf
\theoremstyle{definition}

\usepackage{indentfirst}
    
\usepackage{cleveref} 
  \crefname{section}{Section}{Sections}
  \Crefname{section}{Section}{Sections}
  \crefname{figure}{Fig.}{Figs.}%\figurename
  \Crefname{figure}{Figure}{Figures} 
  \crefname{definition}{Definition}{Definitions}
  \crefname{equation}{}{}
  \Crefname{equation}{Equation}{Equations}
  \crefname{table}{Table}{Tables}
  \Crefname{table}{Table}{Tables}
  
\usepackage{array}
\usepackage{multirow}
\newcolumntype{C}[1]{>{\centering\arraybackslash}p{#1}}

\usepackage{arydshln}

\newcount\Comments  % 1 suppresses notes to selves in text
\newcommand{\kibitz}[2]{\ifnum\Comments=0\textcolor{#1}{#2}\fi}
\definecolor{green}{rgb}{0, 0.65, 0}

\newcommand {\ie} {i.\,e.}
\newcommand {\eg} {e.\,g.}

\newcolumntype{P}[1]{>{\raggedright\arraybackslash}p{#1}}

\begin{document}

\title{Defining Self-adaptive Systems: A Systematic Literature Review
\thanks{This work has been partially funded by the Federal Ministry of Education and Research (BMBF) as part of MANNHEIM-AutoDevSafeOps (01IS22087P).}}

\author{\IEEEauthorblockN{Ana Petrovska}
\IEEEauthorblockA{\textit{Department of Informatics} \\
\textit{Technical University of Munich}\\
ana.petrovska@tum.de}
\and
\IEEEauthorblockN{Guan Erjiage}
\IEEEauthorblockA{\textit{Department of Informatics}\\
\textit{Technical University of Munich} \\
guan.erjiage@tum.de}
\and
\IEEEauthorblockN{Stefan Kugele}
\IEEEauthorblockA{\textit{Technische Hochschule Ingolstadt}\\
\textit{Research Institute AImotion Bavaria} \\
Stefan.Kugele@thi.de}
}

\maketitle

\begin{abstract}
% Background 
In the last two decades, the popularity of self-adaptive systems in the field of software and systems engineering has drastically increased.
However, despite the extensive work on self-adaptive systems, the literature still lacks a common agreement on the definition of these systems.  
To this day, the notion of self-adaptive systems is mainly used intuitively without a precise understanding of the terminology.
Using terminology only by intuition does not suffice, especially in engineering and science, where a more rigorous definition is necessary. 
% Objective
In this paper, we investigate the existing formal definitions of self-adaptive systems and how these systems are characterised across the literature.
Additionally, we analyse and summarise the limitations of the existing formal definitions in order to understand why none of the existing formal definitions is used more broadly by the community. 
% Method
To achieve this, we have conducted a systematic literature review in which we have analysed over 1400 papers related to self-adaptive systems. 
% Results
Concretely, from an initial pool of 1493 papers, we have selected 314 relevant papers, which resulted in nine primary studies whose primary objective was to define self-adaptive systems formally.
% Conclusion
Our systematic review reveals that although there has been an increasing interest in self-adaptive systems over the years, there is a scarcity of efforts to define these systems formally.
Finally, as part of this paper, based on the analysed primary studies, we also elicit requirements and set a foundation for a potential (formal) definition in the future that is accepted by the community on a broader range. 
\end{abstract}

\section{Introduction} \label{sec:intro}
Since the publishing of the famous IBM manifesto on autonomic computing by Kephart and Chess~\cite{Kephart2003} almost two decades ago, the interest in the self-* properties of the systems in software engineering has increased rapidly.
Some of the most broadly spread and often found self-* properties in the literature are: self-adaptation, self-awareness, self-healing and self-organising, just to name a few.
For example, the publications on self-adaptive systems have increased by 304\% in the last twenty years, compared to the 50 years before that (from 1951-2001).\footnote{Source: ACM Digital Library.}

There are many disciplines that have been considering the notion of \emph{adaptation}, for example, biology and evolutionary sciences~\cite{brock2000evolution,zadeh1963definition}, climate change and environmental sciences \cite{moser2010framework,de2009using}, as well as film, cinematography and media studies \cite{hutcheon2012theory}. 
The situation slightly differs in the field of software and systems engineering, where we can observe that the majority of the works available focus only on self-adaptive systems, without tackling and clarifying what is understood under the notion of adaptation in the first place.
Hence, defining the property of \emph{system adaptation} is circumvented by the existing works, although defining what we understand under system adaptation is an essential prerequisite for a subsequent definition of self-adaptive systems.

Suppose we only focus on the available definitions of self-adaptive systems. In that case, we can observe the following: there exist prior works that propose informal definitions of self-adaptive systems as part of their papers~\cite{krupitzer2015survey,de2017software,weyns2019software}.
However, all the informal definitions only rely on intuitive understanding communicated by the spoken language that is fairly ambiguous, which results in an under-specified usage of the terminology of self-adaptive systems. 
In response, to tackle the limitations of the informal definitions, some researchers have put the focus on defining these systems formally~\cite{weyns2012forms,bruni2012conceptual}. 
However, despite the notable advancements in the research on self-adaptive systems in the last two decades and the domain's active community, none of the existing formal definitions is broadly accepted and used as a means of communication among the experts in the field.  
Therefore, the understanding of the core terminology still remains imprecise.
To summarise, there is only an intuitive understanding of self-adaptive systems without a more profound understanding and a precise definition of these systems and how they differ from the ``ordinary'' systems considered non-adaptive. 
Furthermore, defining the property of \emph{system adaptation} is the first step toward defining self-adaptive systems, and this is something that this research field has not paid enough attention to yet. 

Other existing works in the literature also support our observations:
Broy~\cite{broy2009formalizing} and Lints~\cite{lints2012essentials} have independently reached the same concision regarding the intuitive use of the terms of adaptation and self-adaptive systems, arguing that although in some instances such intuitive usage might suffice, this is not the case in engineering and science, where a more rigorous definition is necessary~\cite{lints2012essentials}. 
Additionally, Weyns~\cite{weyns2020introduction} in a recent work states that self-adaptive systems are not defined yet and that the lack of broadly accepted definitions is possibly the biggest challenge in the field of engineering self-adaptive systems~\cite{weyns2017software,weyns2013claims}.

\textbf{Problem.} The lack of precise understanding of \emph{what} self-adaptive systems are has different software engineering consequences and implications, for instance,
\emph{how} to build or engineer these systems that go beyond the famous MAPE-K conceptual model.
The fundamental issue with the MAPE-K is that it serves as a reference model for engineering not only self-adaptive, but any self-* system in general.
Although MAPE-K gives some intuition behind the engineering of self-adaptive systems, primarily by the separation of concerns between the managed system and the managing system, a more specific semantics of these two components within the conceptual model is still lacking.
A more specific semantics accompanying the MAPE-K reference model, will also enable a better separation and characterisation of, \eg, self-adaptive, self-organising and self-aware systems. 

Moreover, as mentioned before, besides the engineering implications, the lack of a concrete definition of self-adaptive systems has various scientific consequences. 
Namely, it hinders constructive scientific debates, which are impossible if experts have different understandings of what self-adaptive systems are. 
A better semantics of self-adaptive systems will 
\begin{enumerate*} [1)]
    \item set a foundation for more constructive scientific debates,
    \item complement the already existing works (methods, architectures, models, etc.) in this field, and
    \item set the foundation on \emph{how} to evaluate and compare these systems in the future.
\end{enumerate*}

\textbf{Gap.} Despite
\begin{enumerate*} [1)]
    \item the acceptance and the acknowledgement of adaptation as an emerging property of software systems, and
    \item the various systematic mapping studies and literature reviews in the field of self-adaptive systems~\cite{muccini2016self,macias2013self,krupitzer2018comparison,weyns2013claims,quin2021decentralized},
\end{enumerate*}
to the best of our knowledge, there is no other study that investigates and summarises how self-adaptive systems have been previously defined and characterised in the literature. 
In particular, no prior work summarises and analyses the existing formal definitions of self-adaptive systems in order to understand and gain insight into why none of these formal efforts is accepted by the community and what are their concrete limitations.   

\textbf{Solution.} As a result, as part of this paper, we conduct a systematic literature review, which aims at summarising and analysing the existing works that formally define and specify self-adaptive systems.
The following central research question leads our research:
\begin{tcolorbox}[breakable, enhanced]
How are self-adaptive systems \emph{formally} defined in the literature?
\end{tcolorbox}
To tackle this broad research question, we derive three more refined research questions (further explained in~\cref{sec:methodology}), investigating 
\begin{enumerate*} [1)]
    \item if the existing formal definitions also formalise the notion of \emph{system adaptation} as part of their contributions,
    \item which characteristics of self-adaptive systems are considered in the existing formalism, and 
    \item the formal notations used in each of the studies. 
\end{enumerate*}

\textbf{Contribution.} Our systematic literature review provides an overview of the current state-of-the-art and structures the existing knowledge on how self-adaptive systems have been defined in the literature so far.
More importantly, we analyse and summarise the limitations of the existing formal definitions, which provide new insights into why none of the formal definitions and specifications is accepted and used more broadly by the community. 
Our contributions also provide a foundation for improving the semantics of the core terminology of self-adaptive systems.
This potentially leads towards a future establishment of a more unified understanding of these systems and, ideally, even to a broadly accepted definition of self-adaptive systems in the near future.
A more profound understanding of the terminology will support the community in setting new challenges and identifying new directions for future research.    

The rest of this paper is structured as follows: \Cref{sec:methodology} presents the methodology used, based on which we conduct our systematic literature review and the identified research questions.
We present the results and answer the research questions in \cref{sec:results}. In \cref{sec:discussion}, we further discuss the main findings, followed by a discussion on the limitations of this review.
In~\cref{sec:related_work}, we present the related work and finally, \cref{sec:conclusion} concludes the paper.
\section{Literature Review Methodology} \label{sec:methodology}
\begin{figure*}[htb]
    \centering%
    \includegraphics[width=\linewidth]{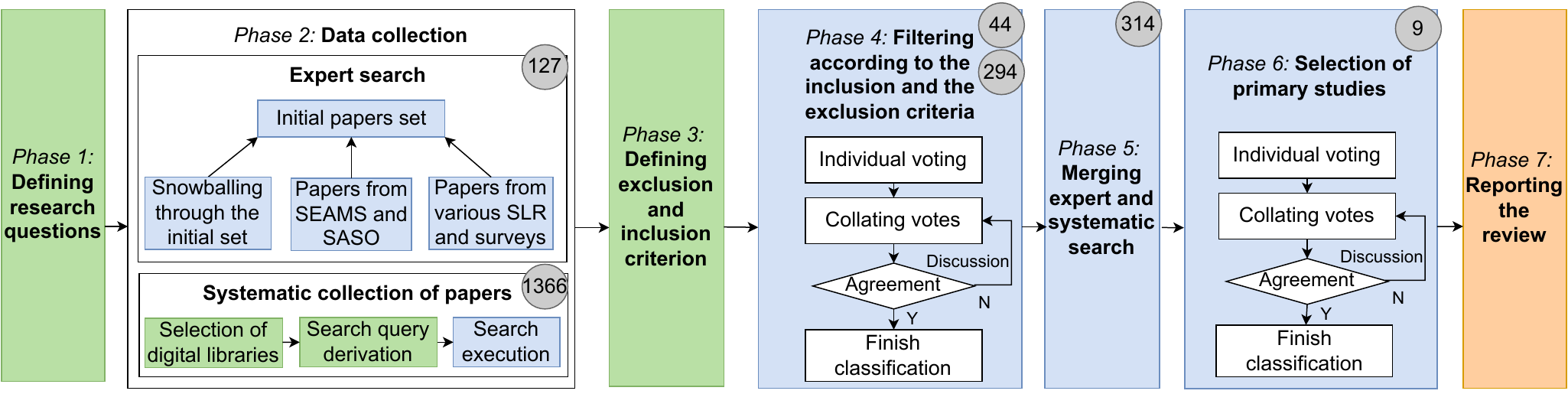}%
    \caption{Research methodology. With green, blue, and orange boxes, we depict the artefacts and the activities in the planning, conducting and reporting of the review. }
    \label{fig:methodology}
    \vspace*{-3mm}
\end{figure*}

This section describes the research methodology we followed in conducting the systematic literature review. 
The systematic process followed the guidelines proposed in various works by Kitchenham~et~al.~\cite{kitchenham2007guidelines,kitchenham2010systematic}.  
An overview of our complete methodology is presented in \cref{fig:methodology}. 

\subsubsection*{\textbf{Phase 1: Defining research questions}} 
The overall objective of the systematic literature review was to give an overview of the current state-of-the-art regarding the definition of self-adaptive systems in software and systems engineering, concretely \textbf{how self-adaptive systems have been \emph{formally} defined in the existing literature}, which is \emph{the leading research question} as part of this work.  
To support answering the leading research question in more detail, we derive three refined research questions:  
%%% Research questions 
\begin{enumerate}[leftmargin=1.05cm,label=\textbf{RQ-\Alph*}]
		\item \label{rq-a} Do the papers with formal definitions of self-adaptive systems also define \emph{system adaptation} as part of their contributions? 
		\item \label{rq-b} Which characteristics of the self-adaptive systems are considered in the existing formal definitions and specifications?
		\item \label{rq-c} Which formal notations have been used across different works to define self-adaptive systems? 
\end{enumerate}

\subsubsection*{\textbf{Phase 2: Data collection}} In this study, we collected the papers in two different ways: manually by an expert search and by conducting a systematic studies collection. 

%\noindent 
\textbf{Expert search.} 
We initially started the data collection by collecting papers in a non-systematic way, referred to as an expert search. 
In the expert search, we started with an initial set of papers that included relevant studies based on our domain knowledge, known to us as key contributions in the field of self-adaptive systems. 
We extended this initial set of studies in three different ways. 
First, we snowballed through the related work of the initial set of studies, as described by Wohlin~\cite{wohlin2014guidelines}. 
Second, we searched through the relevant papers in the conference proceedings of SEAMS\footnote{SEAMS stands for International Symposium on Software Engineering for Adaptive and Self-Managing Systems.} and SASO\footnote{SASO (currently ACSOS) stands for International Conferences on Self-Adaptive and Self-Organizing Systems.} as the two most relevant venues in this domain of research. 
Finally, we searched for relevant papers from previously published systematic literature reviews and surveys on self-adaptive systems. 
In this last step, we considered the studies from Weyns~et~al.~\cite{weyns2012survey}, Muccini~et~al.~\cite{muccini2016self}, Mac{\'\i}as-Escriv{\'a}~et~al.~\cite{macias2013self}, and Krupitzer~et~al.~\cite{krupitzer2015survey,krupitzer2018comparison}. 
At the end, our expert search resulted in 127 relevant studies in total.

%\noindent 
\textbf{Systematic studies collection.} Our systematic search and collection of studies consist of two aspects: 
\begin{enumerate*} [1)]
    \item \emph{selection of digital libraries} on which we perform the automated search, and 
    \item \emph{search query derivation}, which we later used as search queries in the selected databases.   
\end{enumerate*}

We chose the following sources to perform the search:
\begin{itemize} 
    \item ACM Digital Library (\url{https://dl.acm.org/})
    \item IEEE Xplore (\url{http://ieeexplore.ieee.org/})
    \item Scopus (\url{https://dl.acm.org/})
    \item ScienceDirect (\url{http://www.sciencedirect.com/})
    \item Wiley InterScience (\url{http://onlinelibrary.wiley.com/})
    \item World Scientific (\url{https://www.worldscientific.com/})
\end{itemize}

The search query derivation was an iterative process.
Concretely, half a dozen trial searches were performed in each database to evaluate the number of relevant studies obtained by different queries. 
Through this iterative process, we aimed to better understand the suitability of different search queries and keyword combinations, their advantages, and limitations, which was crucial for the final keyword query selection. 
Namely, we aimed for a search query as general as possible to consider a broad range of relevant papers from the literature while minimising the number of irrelevant studies. 
Some of the initial searching queries were the following: \texttt{(self-adapt* AND software)}, \texttt{(self-adapt* AND system)}, and \texttt{(self-adapt* AND engineer*)}.
Our preliminary results showed that including the keywords \texttt{system} and \texttt{engineer*} in the query resulted in many irrelevant studies, \eg\ from networks and hardware. 
On the opposite side, we realised that restricting only to the keyword \texttt{software} excludes works from the domain of cyber-physical systems, which are systems with increasing prominence in the field of self-adaptive systems in the last decade. 
Furthermore, since the main focus of this literature review is to get a better understanding of how self-adaptive systems are defined, we also tried using  \texttt{(self-adapt* AND defin*)} as a searching query, which unfortunately gave only a few results.   
Different combinations of these searching keywords have led either to a broad set of irrelevant papers or to a very narrow search. 
For that reason, we used \texttt{(self-adapt* AND (software OR cyber-physical))} as a final searching query for our automated search on the databases identified above, searching by meta-data (title, abstract, keywords) or only by title, depending on the advanced search options available for the chosen databases. 
The systematic collection resulted in 1366 studies matching the derived searching query.

\subsubsection*{\textbf{Phase 3: Defining inclusion and exclusion criteria}} After the collection of the papers, we needed to perform the first study selection.
Since we are exclusively interested in studies related to system adaptation and engineering self-adaptive systems, in this phase, we defined rigorous inclusion and exclusion criteria to filter the irrelevant papers collected during the extensive search in the previous phase. 
The inclusion and the exclusion criteria that we defined for this purpose are presented in \cref{table:inclusion_criteria} and \cref{table:exclusion_criteria}, respectively.
\begin{table}%[h!]
\caption{Inclusion criteria.}%
\label{table:inclusion_criteria}%
\begin{tabularx}{\columnwidth}{@{}cX@{}}
\toprule
\textbf{Criteria} & \textbf{Description}\\\midrule
 I1   & Papers that have been published in conferences and journals, including full research papers, short papers, position papers, new ideas and emerging results papers, and papers from doctorate symposiums  \\ 
 I2   & Literature published in book chapters  \\ 
 I3   & Papers defining adaptivity, context, self-adaptivity in software engineering \\ 
 I4   & Papers proposing engineering approaches (for example, frameworks, methodologies, methods, reference architectures) for self-adaptive systems \\ 
 I5  & Papers focusing on modelling, design, architecture, and engineering of self-adaptive systems \\ 
 I6   & Systematic literature reviews and mapping studies on self-adaptive systems  \\
 \bottomrule
\end{tabularx}
\end{table}

\subsubsection*{\textbf{Phase 4: Filtering according to the inclusion and exclusion criteria}} In this stage, we apply the inclusion and exclusion criteria to the studies collected through
\begin{enumerate*} [1)]
    \item the expert search, and
    \item the automated systematic collection.  
\end{enumerate*}
Two of the co-authors of this literature review performed the filtering and selection of the studies in this stage. 
During the voting process, the title, the abstract, and, if necessary, the introduction and conclusion of each study (1493 in total: 127 from the expert search and 1366 from the systematic collection) were read and carefully examined to determine their relevance. 
The exact steps of the classification and the voting process of this phase are depicted in \cref{fig:methodology}.
In a nutshell, the authors voted and classified each paper individually.
A discussion followed if the authors' votes were in disagreement until the authors reached a unified decision about the study under analysis.   
Applying the inclusion and the exclusion criteria resulted in 338 studies in total: 44 studies from the expert search and 294 studies from the systematic collection. 
\begin{table}%[h!]
\caption{Exclusion criteria.}%
\label{table:exclusion_criteria}%
\begin{tabularx}{\columnwidth}{@{}cX@{}}  \toprule
\textbf{Criteria} & \textbf{Description}\\\midrule
 E1   & Papers that are not full research papers, including abstracts, tutorials, presentations, or lecture notes  \\ 
 E2   & Papers without PDF and abstracts \\ 
 E3   & Tool papers, case studies, roadmaps, overviews \\ 
 E4   & Papers not focusing on self-adaptivity in software engineering and cyber-physical systems, but instead focus on  \\ 
    & \begin{tabular}{p{0.3cm} p{6.5cm}} 
        E4.1   & Specialised parts of software engineering, for example: software product lines, cloud-based, service-oriented \\ 
        E4.2   & Other computer science fields: networking, hardware, middleware, OS, sensing, control and control theory or static robotics systems \\ 
        E4.3   & Another field in general, for example energy, manufacturing, smart buildings, smart cities, traffic control, economics, natural processes, etc.\\ 
    \end{tabular} \\
 E5   & Papers not focusing on modelling, design, architecture, and engineering of self-adaptive systems, but instead on verification, validation, testing, or operation and maintenance phases of a software life cycle  \\
 E6   & Papers not written in English \\  \bottomrule
\end{tabularx}
\end{table}

\subsubsection*{\textbf{Phase 5: Merging expert and systematic search}} In this phase, the filtered results from both the expert search and the systematic collection from the previous phase are combined, and the found duplicates are removed. 
This resulted in 314 unique and relevant papers. 

\subsubsection*{\textbf{Phase 6: Selection of primary studies}} 
\begin{figure}%[htb]
    \centering%
    \includegraphics[width=0.9\columnwidth]{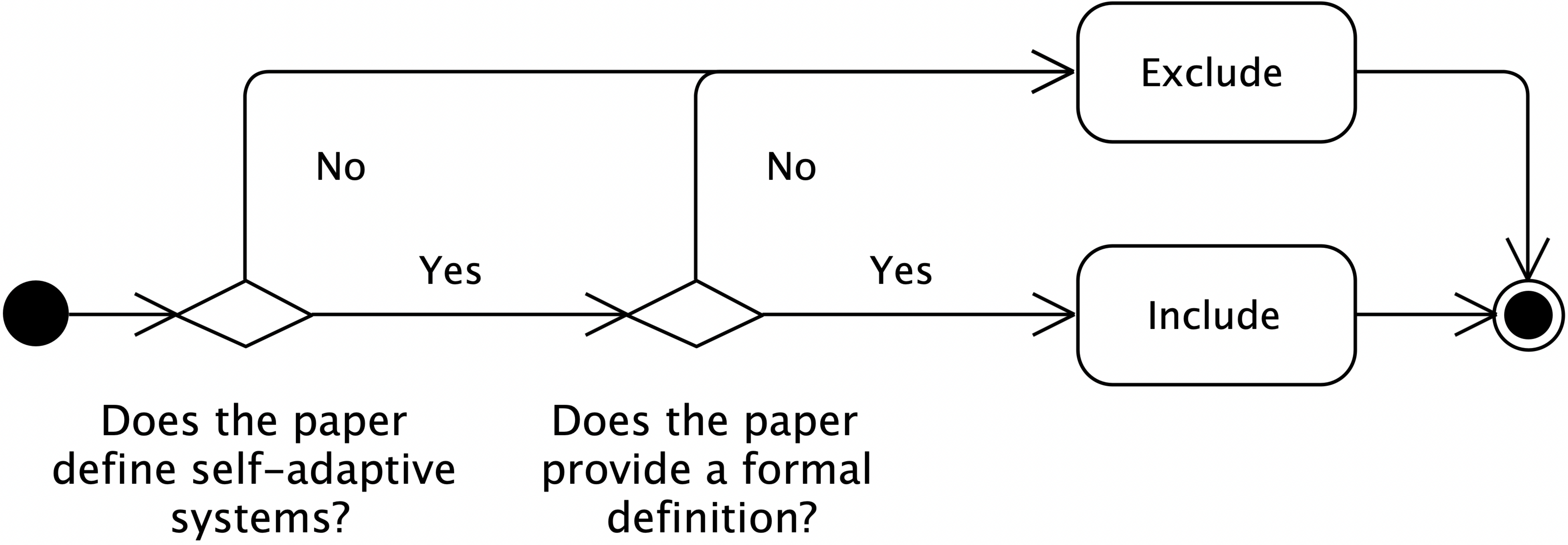}%
    \caption{Two-step selection process.}
    \label{fig:voting_process}
    \vspace*{-4mm}
\end{figure}

The selected relevant papers from the previous phase could be analysed in different ways based on the aims and the goals of the concrete study. 
Since in our work we are interested in how self-adaptive systems are \emph{formally} defined in the literature, we analysed and classified the relevant studies from Phase 5 according to two questions, depicted in the activity diagram in \cref{fig:voting_process}. 
The selection process in this phase was similar to the inclusion and exclusion criteria filtering process described previously in Phase 4. 
In summary, two of the authors independently analysed and classified the 314 relevant studies from the previous step, according to the two-step selection process from~\cref{fig:voting_process}. 
The votes were consolidated, and in case of disagreements, discussions took place among the authors until reaching a unified decision. 
Applying the two-step selection process in this phase resulted in a final set of \emph{nine primary studies} that are analysed rigorously in the rest of this paper. 

Please note the following about our analysis: 
\begin{enumerate*} [1)]
\item there were more than nine studies that included some formalism; however, we only selected those papers whose goal was to define self-adaptive systems formally and the studies with different objectives were excluded during this selection process, and 
\item there were three more papers~\cite{Haesevoets2009, Hachicha2016, Hachicha2018} that claimed to define self-adaptive systems in their abstract and introduction, but since the actual contributions of these papers did not fulfil their claims, we excluded them from our primary studies. 
\end{enumerate*}

\subsubsection*{\textbf{Phase 7: Reporting the review}} A reproducible package with the selected studies in each of the phases of our methodology, the authors' voting, and the analysed data is available online.\footnote{\url{https://github.com/tum-i4/self-adaptive_SLR}}
Additionally, the package contains the BibTeX bibliography (.bib) of all the relevant studies from Phase 5. 
\section{Results} \label{sec:results}

\subsection{General overview of the results } \label{subsec:results_overview}
\begin{figure}[h]
    \centering%
    \includegraphics[width=1.05\linewidth]{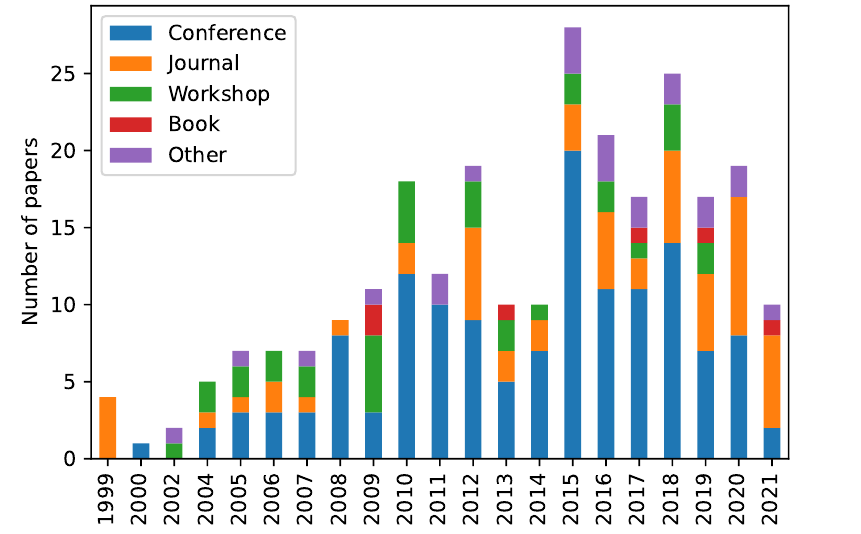}%
    \caption{Overview of the number of publications per year.}
    \label{fig:paper_per_year}
\end{figure}

This section gives an overview of the 314 relevant papers analysed in Phase 6.
In \cref{fig:paper_per_year}, we show the distribution of the papers over the years in different types of venues.
We can also see in~\cref{fig:paper_per_year} that the first works on this topic were published in 1999, and the publication trend has grown since 2004, which can be correlated with two distinct events. 

The first event is related to the first noted instance of the term self-adaptive software in the literature in a technical report by Laddaga in 1997~\cite{laddaga1997self}. 
In this report, Laddaga informally defines self-adaptive software systems as ``[\dots] software that evaluates its own behaviour and changes behaviour when the evaluation indicates that it is not accomplishing what the software is intended to do, or when better functionality or performance is possible.''
The author also adds that the research in self-adaptive systems ``[\dots] seeks a new basis for making software adaptive, that does not require specific adaptive techniques, such as neural networks or genetic programming, but instead relies on software informed about its mission and about its construction and behaviour.'' 
The second, probably even more significant event was the publishing of the famous IBM manifesto on autonomic computing by Kephart and Chess~\cite{Kephart2003} in 2003.
This paper introduced the MAPE-K conceptual model and set the foundation not only for engineering self-adaptive systems but self-* systems in general, \eg, self-awareness~\cite{lewis2011survey,kounev2017notion} and self-healing~\cite{psaier2011survey,ghosh2007self}. 
The manifesto on autonomic computing also set a foundation for a whole new research field on self-adaptive systems, which has been expanding for the last two decades. 

\begin{figure}[h]
    \centering%
    \vspace*{-3mm}
    \includegraphics[width=.8\columnwidth]{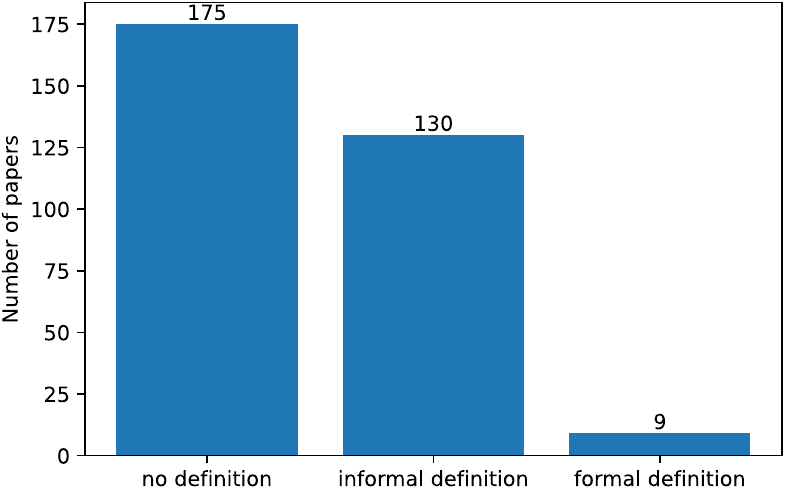}%
    \caption{Overview of the type of definitions.}
    \label{fig:paper_types}
    \vspace*{-4mm}
\end{figure}

\Cref{fig:paper_types} shows that from the 314 relevant studies we analysed in Phase 6, the majority of studies---56\% of the studies (175 papers) provide neither an informal or formal definition nor an intuition of the authors' understanding of self-adaptive systems.
We did not expect these results since these studies were rigorously selected to contribute with solutions for engineering self-adaptive systems.  
41\% of the studies (130 papers) provide informal definitions as part of their works, and \emph{only 3\% of the studies (9 papers) provide some formalisation} of the notion of self-adaptive systems. 
We selected those nine studies as primary for further analysis in our systematic review. 
\subsection{Identifying the different classes and dimensions for analysis} \label{subsec:dimensions-for-analysis}
To answer the leading research question as part of this work and to discuss how self-adaptive systems are formally defined in the literature, we introduce four classes of analysis dimensions: \textbf{(C1)} papers that formally define the property of \emph{system adaptation} as part of their formal definition of self-adaptive systems, \textbf{(C2)} papers that formalise MAPE behaviour, \textbf{(C3)} papers that consider different characteristics of self-adaptive systems in their formal definitions, and \textbf{(C4)} used formal notation.
The introduced analysis classes contain eight analysis dimensions in total, based on which we analysed all the primary studies (see~\cref{table:formal_definitions}). 

As discussed previously in~\cref{sec:intro}, in order to define self-adaptive systems, we first need to understand what the notion of \emph{system adaptation} means in the field of software and systems engineering.
Defining adaptation as a system property is
\begin{enumerate*} [1)]
    \item the core pillar for defining self-adaptive systems, and
    \item is necessary to compare the existing and future works in this field. 
\end{enumerate*}
Therefore, we want to investigate if the existing papers on formalising self-adaptive systems also define system adaptation as part of their contributions.
Hence, in our first class (C1), we differentiate between 
\begin{enumerate*} [1)]
    \item papers with a concrete aim to \emph{explicitly} formalise system adaptation, 
    \item papers that assume they define this notion \emph{implicitly}, for instance, through formalising adaptive system behaviour, and 
    \item papers that \emph{do not formally define} system adaptation in their work.  
\end{enumerate*}

During our analysis, we also identified that some of the primary studies aimed at defining adaptive behaviour by specifying the behaviour of the \textbf{MAPE-K} feedback loop.
In response, we introduced the second class (C2) for analysis. 

In the third class (C3) of the analysis dimensions, we consider various characteristics identified in the literature as essential while defining self-adaptive systems based on the external and internal principles proposed in a recent work by Weyns~\cite{weyns2020introduction}.
As we elaborated previously, Weyns has stated that there is no consensus on the definition of self-adaptation so far in the community.
In response to that, as part of his work~\cite{weyns2020introduction}, he proposes two complementary \emph{principles}---external and internal---that characterise self-adaptive systems.
The principles are built upon the consolidated usage of the notion of self-adaptive systems for the past decade in the community. 
To the best of our knowledge, this is the most complete consolidated characterisation of self-adaptive systems. For that reason, we used the characteristics from the principles to identify further dimensions for the analysis of our primary studies. 

According to the external principle, a self-adaptive system handles changes and \textbf{uncertainties} from its environment (also referred to as \textbf{context}), the \textbf{system}, and the system goals autonomously. 
The context is the part of the environment relevant to a particular system~\cite{broy2009formalizing}.
These two terms have often been used interchangeably in the domain of self-adaptive systems; however, from our point of view, having a clear understanding and differentiation of context and environment is important.
However, making this differentiation is not the aim of this work, and through the rest of the paper and in~\cref{table:formal_definitions} we will use the concept of \emph{context} only. 
By \emph{system} in the table, we refer to the managed system that gains the ability to adapt as part of a self-adaptive system and not the self-adaptive system as a whole. 

The internal principle separates the system goals in self-adaptive systems into domain and adaptation goals. 
The \textbf{domain goals} are related to the concerns of the managed system---the system that gains adaptation capabilities.
Whereas the \textbf{adaptation goals} are related to the concerns of the managing system---the entity of the self-adaptive system that enabled the adaptation of the managed system.
We use this differentiation from the principles and make a further semantic distinction between the managed and the managing system as part of a self-adaptive system. 
Concretely, we say that the domain goals are related to the functionality of the system---more precisely, to the fulfilment of the system function, \ie, the function of the managed system.
In contrast to the domain goals, we consider the adaptation goals to be in relation to one or more quality criteria or objectives, which is also supported by prior works~\cite{villegas2011framework,weyns2012claims}. 
In sum, we consider the separation of the goals in self-adaptive systems between domain and adaptation goals as essential, as it provides the basis for discussing and distinguishing when a system adapts and when it simply operates or functions. 
Furthermore, during the analysis of our primary studies, we did not concretely search whether these terms are used or not. Instead, we analysed the studies more thoroughly to answer if perhaps the studies adopt these ideas while using different terminology, or if these ideas are implicitly considered in their contributions and formulas without giving them a specific name.   

In a nutshell, in the third class (C3) of analysis dimensions, we differentiate between 
\begin{enumerate*} [1)]
    \item papers that include a concrete characteristic \emph{formally} as part of their definition, 
    \item papers that identify or mention a concrete characteristic only \emph{informally} and do not include it as part of their formalism, and 
    \item papers that \emph{do not} even identify or mention the necessity for the consideration of a concrete characteristic in their definition of self-adaptive systems.  
\end{enumerate*}

Finally, in the fourth class (C4), we have noted the \textbf{formal notation} used in each paper. 

\subsection{Analysis of the primary studies} \label{subsec:primary_studies}
In this section, we analyse the primary studies in order to consolidate the existing work and answer the research questions.
A thorough analysis of the primary studies and discussion of their limitations should enable us to set the foundation for improving the semantics and to derive requirements for a unified and precise definition of self-adaptive systems in the future. 
Although we collected the papers systematically, we ended up only with nine primary studies for the analysis. 
Therefore, we decided to take a more qualitative approach to analyse our primary studies guided by our leading research question that we previously introduced in~\cref{sec:methodology}).
We summarise the qualitative analysis of our primary studies in the following, based on which~\cref{table:formal_definitions} is filled. 
Due to space limitations, we are not giving the formal details, but we include the used formal notations in each of the primary studies. 

One of the first efforts to formally define adaptive behaviour was made by Zhang and Cheng~\cite{Zhang2006Modelbaseddevelopment}, in which the authors proposed a model-driven software development process for dynamically adaptive programs. 
According to the authors, adaptive programs are generally more difficult to specify due to their high complexity, especially in multi-threaded adaptations where the program behaviour results from the collaborative behaviour of multiple threads.
This is the first main limitation of this work since adaptation is not necessarily an emerging property from a collaboration, and it should be treated and defined as a separate concept.
In their formal representation of adaptive programs, a \emph{program} is represented by a state machine that exhibits certain behaviour and operates in specific domains. 
A dynamically adaptive program operates in different domains and changes its behaviour (\ie, behavioural modes corresponding to the specific domain) at run-time in response to domain changes. 
As part of their work, the authors do not explicitly formalise system adaptation; however, they illustrate the specification process for three types of adaptive behaviour by modelling an audio streaming protocol with Petri nets. 
The authors use prior works on specifying dynamic systems architectures~\cite{allen1998specifying,bradbury2004survey} to formalise adaptive programs.
As a result, they often use the terms adaptive and dynamic interchangeably throughout the paper without clearly distinguishing between them, which is the second limitation of this study.
Lastly, this work does not consider any of the other analysis dimensions identified in class C3, which are paramount to be included in a holistic formal definition.   

A similar concept in which adaptation is described through the realisation of different behavioural modes is proposed by Klarl~\cite{Klarl2015}.
In this work, the author realises the behavioural modes by roles which can be dynamically adopted by a component. 
Concretely, the author proposes a model-driven engineering process to develop self-adaptive systems, in which the adaptation logic (\ie, the managing system) is considered independently from the application logic (\ie, the managed system) and supports the systematic transition between their components. 
For specification, the author proposes hierarchical adaptation automata, and for the design---a role-based architecture according to a Helena Adaptation Manager pattern.
This study neither defines the notion of system adaptation nor adaptive behaviour.
Except for considering the context (concretely, perceptions about the context) and the system state as attributes of the signature of self-adaptive component types in the formalism of the paper, no other analysis dimension from class C3 is considered as part of this study.  

In two separate works, Broy~et~al.~\cite{broy2009formalizing} and Bruni~et~al.~\cite{bruni2012conceptual} try to answer how the self-adaptive systems differ from the ``ordinary'' systems, which are considered non-adaptive. 
Concretely, Broy~et~al. aim at defining adaptive system behaviour while differentiating interaction patterns between three separate entities: the system, a subject (a user or other technical system that interacts with the system) and the context.
The authors claim that one can differentiate the adaptive behaviour of the system only by considering and observing the context in which the system operates. 
The authors further classify the system inputs into direct/explicit and indirect/implicit, and assume that a system always receives the user inputs explicitly.
Therefore, adaptive system behaviour can be observed if the system reaction resulting from the user input (the explicit input) is additionally determined by some additional information about the context received through the implicit inputs. 
Based on these ideas, the authors identify four types of observable system behaviour (\ie, adaptive behaviour) with respect to the user: non-adaptive, non-transparent adaptive, transparent adaptive and diverted adaptive behaviour.
To summarise, as part of this work, the authors identify the consideration of the context and the system (state) as relevant and necessary for the system adaptation; and therefore, they include them as part of their formalism, which is based on \textsc{Focus} modelling approach.

\renewcommand\tabularxcolumn[1]{m{#1}}% 
\begin{table*}
\centering
\caption{Summary of papers that provide some formal definitions on system adaptation and self-adaptive systems.}%
\label{table:formal_definitions}%
\begin{tabularx}{\textwidth}{@{}l@{\hskip 2mm}l@{\hskip 2mm}XXXXXXXXXX@{}}
\toprule
\textbf{Class} & \textbf{Analysis Dimension} & 
Zhang \& Cheng~\cite{Zhang2006Modelbaseddevelopment}, \citeyear{Zhang2006Modelbaseddevelopment} &  \citeauthor{broy2009formalizing}~\cite{broy2009formalizing}, \citeyear{broy2009formalizing}  & \citeauthor{Qureshi2011}~\cite{Qureshi2011}, \citeyear{Qureshi2011} & \citeauthor{bruni2012conceptual}~\cite{bruni2012conceptual}, \citeyear{bruni2012conceptual} & \citeauthor{weyns2012forms}~\cite{weyns2012forms}, \citeyear{weyns2012forms}  & \citeauthor{Arcaini2015a}~\cite{Arcaini2015a}, \citeyear{Arcaini2015a} & 
\citeauthor{Iglesia2015}~\cite{Iglesia2015}, \citeyear{Iglesia2015} & \citeauthor{Klarl2015}~\cite{Klarl2015}, \citeyear{Klarl2015} &  Bucchia\-rone and Mongiello \cite{bucchiarone2019ten}, \citeyear{bucchiarone2019ten}  \\
\midrule
\textbf{C1} & \textbf{System adaptation} & Implicit & Implicit & No & No & No & No & No & No & Explicit\\ 
\midrule
\textbf{C2} & \textbf{MAPE behaviour} & No & No & No & No & No & Yes & Yes & No & No\\ 
\midrule
& \textbf{Uncertainties} & No & No & No & No & No & No & No & No & No\\
& \textbf{Context (state)} & No & Formal  & Formal & Formal & Formal & Formal & Formal & Informal & Formal\\ 
\textbf{C3} & \textbf{System (state)} & No & Formal & No & Formal & Formal & Formal & Formal & Formal & Formal\\ 
& \textbf{Domain goals} & No & No & Formal & No & No & No & No & No & No \\
& \textbf{Adaptation goals} & No & No & Formal & No & Informal & Informal & Formal & No & No\\\midrule
\textbf{C4} & \textbf{Formal notation} & Petri nets & FOCUS & Techne & LTS & Z language & ASM & TA, TCTL & LTS & TGG, LTS\\ \bottomrule\\[-1mm]
\end{tabularx}
\vspace*{-6mm}
\end{table*}
Bruni~et~al.~\cite{bruni2012conceptual} propose a conceptual framework for adaptation, in which they assign a central role to control data, which governs the adaptive behaviour of a component. 
The authors define adaptation informally as a run-time modification of the control data and, consequently, consider a component as self-adaptive if it can modify its own control data at run-time. 
They formally define adaptable vs non-adaptable components, self-adaptive components, and knowledge-based adaptation, in which they recognise the context as the observable part of the environment.
The authors formalise their conceptual framework using a Labelled Transition System (LTS) model. 
Similarly as in~\cite{broy2009formalizing}, the authors consider the context and the system state as part of their formalisation; however, all the other analysis dimensions from class C3 are not considered in either of these two works.
The most significant shortcoming of this work is that the central idea of their concept (\ie, the control data) is left fuzzy and unclear since the authors do not elaborate precisely on what they understand under the notion of control data, how one can identify control data in the system, how the system is influenced by the control data and the structure of the control data. 
Furthermore, in contrast to the work by Broy~et~al.~\cite{broy2009formalizing}, Bruni~et~al. do not formalise or specify adaptive behaviour as part of their work. 

Weyns~et~al.~\cite{weyns2012forms} propose formally specified models for designing self-adaptive software systems. 
The authors propose a FOrmal Reference Model for Self-adaptation (FORMS), which enables precise descriptions of the architectural characteristics of distributed self-adaptive software systems in the early design phases of the system.
FORMS primarily focuses on the formalisation of the structural aspect of self-adaptive systems without providing any insights into the behavioural semantics of the self-adaptive systems. 
Although FORMS had and continues to have a notable impact in the community, it neither defines system adaptation nor adaptive behaviour. 
FORMS considers the aspect of context and system formally, and the adaptation goals are only considered informally throughout the work.    
Finally, similarly to~\cite{Zhang2006Modelbaseddevelopment}, the authors of FORMS leverage some other concept---specifically in FORMS, the notion of system distribution---to compensate in some sense for the lack of precise understanding of system adaptation necessary for the definition of self-adaptive systems.

Arcaini~et~al. in~\cite{Arcaini2015a} and Inglesia and Weyns in~\cite{Iglesia2015} aim to define self-adaptive systems by formally specifying the MAPE-K feedback loop.
Arcaini~et~al.~\cite{Arcaini2015a} show how MAPE-K loops can be explicitly formalised in terms of agents' actions using Abstract State Machines (ASM) transition rules to model the behaviour of self-adaptive systems. 
Concretely, the authors exploit the concept of multi-agent ASM to specify decentralised adaptation control by using MAPE computations. 
Although the authors aim at modelling and specifying self-adaptive systems, concretely the behavioural aspect of self-adaptation, their contribution primarily focuses on specifying the behaviour of the MAPE feedback loop (\ie, the managing system) and not the behaviour of the self-adaptive system as a whole.
The other shortcoming is that the authors consider the adaptation as a result of the collaborative behaviour of multiple managing agents (\ie, MAPE-K loops). However, system adaptation is not necessarily an emerging property from collaboration, and its definition should be independent of the type and nature of the system.
Finally, the authors consider the context and system in their formal specifications and informally the adaptation goals.  

To support the design and the engineering of self-adaptive systems, Inglesia and Weyns in~\cite{Iglesia2015} derive a set of MAPE-K formal templates for designing feedback loops of self-adaptive systems.
The proposed templates comprise:
\begin{enumerate*} [1)]
    \item behaviour specification templates for modelling different components of the MAPE-K loop and their interaction---using networks of timed automata (TA), and 
    \item and property specification templates for specifying required properties of the adaptive behaviour---based on timed computation tree logic (TCTL). 
\end{enumerate*}
Similar to the work of Arcaini~et~al.~\cite{Arcaini2015a}, the authors of~\cite{Iglesia2015} do not define the adaptive behaviour of the entire self-adaptive system but instead specify the MAPE behaviour, assuming that the MAPE behaviour will eventually adapt the managed system. 
As part of this work, the context, the system, and the adaptation goals are formally considered in the templates. 

A more complete formalism has been proposed in a recent work by Bucchiarone and Mongiello~\cite{bucchiarone2019ten}, in which the authors introduce a formal framework to characterise different aspects of an ensemble-based software engineering. 
Concretely, they present
\begin{enumerate*} [1)]
\item how to model dynamic software ensembles using Typed Graph Grammar (TGG),
\item how to specialise and re-configure ensembles, and 
\item how to manage collective adaptations in an ensemble.
\end{enumerate*}
As part of this work, the authors use TGGs combined with Labelled Transition Systems (LTSs) to formally define system context, context-awareness, and system adaptation; however, only in the frame of system ensembles, which is the biggest shortcoming of this paper.
However, adaptation as a system property should be considered and defined in independence from ensembles or system collaboration and not as an emerging property thereof. 
It is important to point out that compared to all the other analysed primary studies, there is a notable maturity in the work by Bucchiarone and Mongiello~\cite{bucchiarone2019ten}.
Concretely, this is the only work that defines system adaptation as part of their contribution formally.  
Furthermore, the authors also identify the importance of considering the context and the system, by explicitly considering the system functionality that adapts, as necessary aspects to discuss system adaptation and, therefore, self-adaptive systems.  

Qureshi~et~al.~in~\cite{Qureshi2011} take a different approach than the rest of the primary studies.
In their work, the authors focus on defining the requirements for self-adaptive systems instead of defining self-adaptive systems.
The authors tackle how the requirements problems (\ie, the problems solved during the requirements engineering) differ for self-adaptive systems compared to systems that are not self-adaptive.
As it was previously observed, Broy~et~al.~\cite{broy2009formalizing} and Bruni~et~al.~\cite{bruni2012conceptual} also tried to differentiate in their works how self-adaptive systems differ from those that are considered non-adaptive. 
The overarching objective of the work by Qureshi~et~al.~in~\cite{Qureshi2011} is to identify concepts and relations that are necessary to be considered while eliciting and analysing requirements for self-adaptive systems. 
Therefore, the authors do not aim to define system adaptation, adaptive behaviour, nor MAPE behaviour as part of their work.
Although this paper does not explicitly identify the relevance of the independent consideration of the system (\ie, the managed system that gains adaptation capabilities) as part of their formalism, this is the only paper in our primary studies that makes a distinction and formally considers the domain goals (referred to as mandatory goals as part of their work), and the adaptation goals (referred to as quality constraints). 

\subsubsection*{\textbf{Addressing RQ-A}} \emph{Do the papers with formal definitions of self-adaptive systems also define \emph{system adaptation} as part of their contributions? }
It is not possible to define self-adaptive systems without defining what it means for a system to adapt in the first place.
However, our literature analysis showed that only one study formally defines system adaptation as part of their efforts to define self-adaptive systems; however, only in the frame of system ensembles. 
Two primary studies implicitly define system adaptation by specifying adaptive system behaviour as part of their contributions.
And finally, two studies specify the MAPE behaviour (\ie, the behaviour of the managing system as part of a self-adaptive system), assuming that the MAPE behaviour will eventually adapt the managed system.

\subsubsection*{\textbf{Addressing RQ-B}} \emph{Which characteristics of the self-adaptive systems are considered in the existing formal definitions and specifications?}
If in RQ-A we focused on the behavioural aspect of self-adaptive systems, in RQ-B, we shift the focus to the structural aspects of these systems.
Concretely in this research question, we investigate which of the characteristics that have been recently consolidated in this field of research, as explained in~\cref{subsec:dimensions-for-analysis}, are considered in the existing body of work that formally defines self-adaptive systems.   
The most notable insight of our analysis is that \emph{none} of the primary studies consider the aspect of uncertainty, both formally and informally, as part of their contribution.
This is extremely surprising since the notion of uncertainty has been at the centre of the idea behind self-adaptive systems.  
Precisely the core motivation for self-adaptive systems is built on the unpredictable changes and \emph{uncertainties} that trigger the need for system adaptation during the run-time of the system. 
This is also roughly how all the informal definitions available in the literature define self-adaptive systems, with a liberate use of the notion of uncertainties---a notion that is seemingly difficult to be put in formalism, as shown by our results. 
These results are another proof of the importance of having a clear, systematic, and formal definition of self-adaptive systems. 

Almost all of the primary studies that we analysed consider the (states of the) context and system in some way as part of their formalism---the majority of them formally.  
This concludes that system adaptation and, therefore, self-adaptive systems cannot be defined in isolation from the context in which the self-adaptive systems operate and the properties of the system (\ie, the managed system) that gains the ability to adapt as part of a self-adaptive system.

Four out of nine primary studies (two formally and two informally) consider the concept of the adaptation goals, as we previously described them in~\cref{subsec:dimensions-for-analysis}, and identify that the system self-adapts in order to fulfil some quality objectives.
However, the number of primary studies that consider the domain goals is much lower, and out of the nine primary studies only one study considers the domain goals.
This is probably because this differentiation and the identification of the domain goals is much more subtle, but as we discussed in~\cref{subsec:dimensions-for-analysis}, it is necessary in order to argue when the system adapts and when does it simply function.

\subsubsection*{\textbf{Addressing RQ-C}} \emph{Which formal notations have been used across different works to define self-adaptive systems?}
Among the primary studies, three papers used Labelled Transition Systems (LTS)---in which one of them used Typed Graph Grammars (TGG) in combination with LTS.
The rest of the studies used: Petri nets, FOCUS, Techne, Z language, abstract state machines (ASM), timed automata (TA) and timed computational tree logic (TCTL).
\section{Discussion} \label{sec:discussion}
\subsection{Discussion on the results and future works}  \label{subsec:discussion-results}
Despite the vibrant and growing community and the expanding interest in self-adaptive systems, our results have shown a sparsity of contributions that define self-adaptive systems formally. 
We derive various premises from the analysis and the results of our study, which set the foundation for the requirements for a holistic, formal definition. 

The ideas of autonomic systems that introduced the \mbox{MAPE-K} conceptual model have profoundly impacted the engineering field and have initiated various new lines of research for the last two decades. 
Although MAPE-K gives some intuition behind the engineering of self-adaptive (and self-*) systems by the separation of concerns between the managing and the managed system, a more specific semantics of these two components is still missing. 
For instance, one can assume that every system that does some monitoring, planning, analysis, and execution and has some loose interpretation of the knowledge (\eg\, every cyber-physical system), is self-adaptive by default.
In response, the principles proposed by~Weyns~\cite{weyns2020introduction}, concretely the internal principle that differentiates between the domain and the adaptation goals, have already made the initial steps in the direction of improving the terminology. 

As we previously discussed in~\cref{subsec:dimensions-for-analysis}, it is paramount to distinguish between system functioning and system adapting. 
Making this distinction will set the foundation for defining \emph{system adaptation} and, subsequently, self-adaptive systems.
In our analysis, we observed that in three of the primary studies~\cite{broy2009formalizing,bruni2012conceptual,Qureshi2011}, the authors raised the question of the necessity to differ \mbox{(self-)}adaptive systems from the ``ordinary'', non-adaptive systems. 
However, the work by Bucchiarone and Mongiello~\cite{bucchiarone2019ten} is the only study that contributes in this direction, in which the authors focus on identifying the system functionality that adapts; therefore, explicitly separating between system functioning and system adapting. 

It is notable from the surveyed literature and our analysis that none of our primary studies (see \cref{table:formal_definitions}) considers all the characteristics of self-adaptive systems as discussed in the principles~\cite{weyns2020introduction}. 
The most unexpected insight from our results is that the notion of uncertainty has not been considered in the contributions of any of the primary studies, although uncertainty is considered the main reason for self-adaptive systems in the published papers on this topic and the informal definitions of these systems. 
So far, there is an intuitive understanding of the concept of uncertainty in self-adaptive systems, resulting in a clear need for more careful consideration of the aspect of uncertainty in this research domain. 
Concretely, how uncertainties can be represented, quantified and in general formalised as part of a formal definition of self-adaptive systems.  

Finally, our results have shown that almost half of the primary studies provide their formalism by leveraging the aspects of collaboration~\cite{Zhang2006Modelbaseddevelopment}, distribution~\cite{weyns2012forms}, decentralisation~\cite{Arcaini2015a}, and ensembles~\cite{bucchiarone2019ten} to define self-adaptive systems.
However, system adaptation is not necessarily an emerging property from collaboration or decentralisation and should be defined independently from these notions.

Based on our results and our findings, we can summarise that a potential formal definition of self-adaptive systems should provide a more precise semantics by 
\begin{enumerate*} [1)]
\item defining what it means for a system to adapt and how system adaptation differs from system function,
\item considering more systematically all the different characteristics of self-adaptive systems in its formalism, in particular the aspect of uncertainty, and
\item defining adaptation and self-adaptive system isolated from, \eg, collaboration and multi-agent systems. 
\end{enumerate*}

\subsection{Threats to validity} 
Although the systematic process for data collection and analysis followed the well-known accepted guidelines for systematic literature review~\cite{kitchenham2007guidelines,kitchenham2010systematic}, there are some possible threats to validity that we summarise in the following. 
\subsubsection*{Internal validity} 
In this study, we aimed to investigate how self-adaptive systems are defined in the literature.
Finding this information in the papers we analysed was not always straightforward, especially while searching for informal definitions since this information was often implicitly included in the text. 
The expertise of the researchers also plays a role in this process; however, the potential bias of the researchers who conduct the systematic literature review is a common threat to validity.
To mitigate this issue, voting was done by two of the authors.
In case of conflicts, there was a follow-up discussion and a more in-depth paper analysis until a consensus was reached. 
On the other hand, searching for the formal definitions in the studies was much less complicated. 
Namely, in this case, we first searched if the analysed studies contained any formalism (which drastically reduced the search space). In case they did, we then proceeded with a thorough analysis of the paper, searching if the paper aims to define self-adaptive systems as part of their (formal) contributions. 
The voting on the formal definitions led to almost no conflicts among the authors. 
\subsubsection*{External validity} Doing an automated search in six databases using the term ``self-adaptive systems'' yields hundreds of thousands of results. 
For that reason, we adopted the following two strategies, as previously explained in \cref{sec:methodology}:
\begin{enumerate}
    \item We implemented an iterative search process with pilot searches to define and fine-tune the search string to minimise the number of irrelevant studies. In each iteration, two authors manually inspected and analysed a subset of the collected data. The search string was refined based on the insights gained from the concrete iteration. 
    \item In our automated search, we either searched in the databases by metadata (title, abstract, keywords) or only by title, depending on the advanced search options available in the concrete database.
    We assumed that if a paper defines self-adaptive systems, then that paper will certainly contain the word \texttt{(self-adapt*)} as part of these fields. 
    However, there is the possibility of having missed some relevant studies by limiting the automated search in the databases only by metadata.
\end{enumerate}
However, not to compromise the completeness of the collected data, we analysed the complete initial pool of papers (1493 papers) and not only a random selection of these works.
This proved to be the right decision, considering that the final set of primary studies contained only nine papers that could have been easily missed if we had decided to analyse only a random selection of the initial pool of papers. 
\subsubsection*{Reliability} To ensure that our research findings can be replicated, as part of this paper, we have made available a reproducible package with the selected studies in each of the phases of the methodology. 
The package contains all the necessary data for replication, including the final queries that we used for the automated search in the databases and the authors' votes.
To mitigate the inherent bias that each researcher has due to their background and experience, we have ensured that multiple researchers made the paper selection and the data extraction and analysis. 
Precisely, during the analysis in Phases 4 and 6 of our methodology, we introduced a voting process in which, if the authors classify a paper differently, a discussion took place until the voters have reached a unified decision about the study. 
On the other hand, the reliability of the used databases and the replication of the automated search with the specific queries is something that we cannot account for. 

\section{Related Work} \label{sec:related_work}
To the best of our knowledge, this is the first study that has focused on systematically collecting and analysing how self-adaptive systems are defined. 
Although the interest in self-adaptive systems has been rapidly growing, the concrete semantics of the core terminology is still missing.
Namely, the literature still lacks a consensus on a definition---understanding why this is the case and getting a better overview of the existing body of literature was the motivating factor for our study. 

Many other systematic reviews and mapping studies with different objectives were conducted over the years in the literature. 
However, they were all focusing on other aspects related to self-adaptive systems, for example, engineering approaches for self-adaptive systems~\cite{krupitzer2015survey,krupitzer2018comparison,weyns2012claims,weyns2013claims}, the use of formal methods in self-adaptive systems~\cite{weyns2012survey}, and two recent works in which the authors focused on decentralisation in self-adaptation~\cite{quin2021decentralized} and the application of machine learning in self-adaptive systems~\cite{gheibi2021applying}.
Besides the existing systematic literature reviews and mapping studies, there are a couple of other surveys and roadmaps on future research challenges~\cite{lemos2013software,macias2013self,salehie2009self,weyns2017software}.
In contrast to these works, in our systematic literature review, we aim to consolidate the existing (formal) definitions of self-adaptive systems and, more importantly, understand their limitations, which sets the foundation for a future establishment of a more unified understanding of these systems. 
An improved terminology semantics will complement the existing works in this field, including the contributions from the other systematic reviews, mapping studies, and roadmaps presented above. 

Motivated by similar incentives as our study, only putting the focus on self-awareness instead of self-adaptation, Elhabbash~et~al.~in~\cite{elhabbash2019self} have conducted a systematic literature review on the usage of self-awareness in software engineering. 
Among other objectives, the authors also summarise and analyse how self-aware systems have been defined in the literature.
Please note that in this study, the authors only focus on summarising the informal definitions of these systems. 
Although most of the researchers in the literature use the terms self-adaptation and self-awareness interchangeably, there are some prior works~\cite{kounev2014descartes,lewis2011survey,petrovska2021self}, in which the authors distinguish these terms and consider self-awareness as an ``enabler'' or a precondition for self-adaptive systems.
In the future, if we have a clearer and more precise definition and understanding of self-adaptive systems, this will also help us to better distinguish self-adaptive systems from other self-* systems, such as self-aware systems. 
\vspace*{-3mm}
\section{Conclusion} \label{sec:conclusion}
The lack of commonly accepted definitions and the ambiguous description of the terminology adds complexity to an already complex field of research. 
In response, our main objective in this paper was to get better insights and summarise the existing body of literature by systematically reviewing how self-adaptive systems have been defined in the prior works in this field.
First and foremost, we wanted to investigate if the existing works define self-adaptive systems and how many of these works provide any definition, putting a special emphasis and qualitatively analysing the formal definitions of these systems later in the paper. 
Our results showed that 1) the majority of the papers we analysed do not even provide an informal definition or description of what they consider a self-adaptive system to be, and 2) only nine papers aimed to define or specify self-adaptive systems in their contributions formally. 
The results of our study clearly revealed that the problem of defining self-adaptive systems remains under-researched, which, to some extent, explains why the research field lacks more unified definitions---or generally a better understanding of the terminology.
We hope that with our study, we raise awareness on this important matter and show the great potential for future research that will focus on creating a more precise understanding and, ideally, formal definitions of self-adaptive systems that will be accepted on a broader range.

Our systematic literature review also provides a foundation for improving the terminology.
Concretely, through our qualitative analysis of the existing formal definitions of self-adaptive systems, we also summarised their limitations and shortcomings, based on which we elicit requirements for a potential holistic and unified formal definition of these systems. 
Increasing the clarity of the terminology will support and complement the already existing approaches and methods for engineering self-adaptive systems and also open new directions of research in the future, enabling the community to endeavour to a fuller extent. 

\section*{Acknowledgment}
We thank Prof. Alexander Pretschner for always asking the hard but meaningful questions, which motivated the idea for this research. Additionally, we thank Prof. Danny Weyns for the valuable feedback on the early draft of this paper.
And finally, we want to thank Thomas Hutzelmann for the continuous support.

\balance
\bibliography{acsos-2022}

\begin{thebibliography}{48}
\providecommand{\natexlab}[1]{#1}
\providecommand{\url}[1]{\texttt{#1}}
\expandafter\ifx\csname urlstyle\endcsname\relax
  \providecommand{\doi}[1]{doi: #1}\else
  \providecommand{\doi}{doi: \begingroup \urlstyle{rm}\Url}\fi

\bibitem[Kephart and Chess(2003)]{Kephart2003}
Jeffrey~O. Kephart and David~M. Chess.
\newblock {The Vision of Autonomic Computing}.
\newblock \emph{Computer}, 36\penalty0 (January):\penalty0 41--50, 2003.
\newblock \doi{10.1046/j.1365-2745.2002.00730.x}.

\bibitem[Brock(2000)]{brock2000evolution}
James~Patrick Brock.
\newblock \emph{The evolution of adaptive systems: The general theory of
  evolution}.
\newblock Elsevier, 2000.

\bibitem[Zadeh(1963)]{zadeh1963definition}
Lotfi~A Zadeh.
\newblock On the definition of adaptivity.
\newblock \emph{Proceedings of the IEEE}, 51\penalty0 (3):\penalty0 469--470,
  1963.

\bibitem[Moser and Ekstrom(2010)]{moser2010framework}
Susanne~C Moser and Julia~A Ekstrom.
\newblock A framework to diagnose barriers to climate change adaptation.
\newblock \emph{Proceedings of the national academy of sciences}, 107\penalty0
  (51):\penalty0 22026--22031, 2010.

\bibitem[de~Fran{\c{c}}a~Doria et~al.(2009)de~Fran{\c{c}}a~Doria, Boyd,
  Tompkins, and Adger]{de2009using}
Miguel de~Fran{\c{c}}a~Doria, Emily Boyd, Emma~L Tompkins, and W~Neil Adger.
\newblock Using expert elicitation to define successful adaptation to climate
  change.
\newblock \emph{Environmental Science \& Policy}, 12\penalty0 (7):\penalty0
  810--819, 2009.

\bibitem[Hutcheon(2012)]{hutcheon2012theory}
Linda Hutcheon.
\newblock \emph{A theory of adaptation}.
\newblock Routledge, 2012.

\bibitem[Krupitzer et~al.(2015)Krupitzer, Roth, VanSyckel, Schiele, and
  Becker]{krupitzer2015survey}
Christian Krupitzer, Felix~Maximilian Roth, Sebastian VanSyckel, Gregor
  Schiele, and Christian Becker.
\newblock A survey on engineering approaches for self-adaptive systems.
\newblock \emph{Pervasive and Mobile Computing}, 17:\penalty0 184--206, 2015.

\bibitem[De~Lemos et~al.(2017)De~Lemos, Garlan, Ghezzi, et~al.]{de2017software}
Rog{\'e}rio De~Lemos, David Garlan, Carlo Ghezzi, et~al.
\newblock Software engineering for self-adaptive systems: Research challenges
  in the provision of assurances.
\newblock In \emph{Software Engineering for Self-Adaptive Systems III.
  Assurances}, pages 3--30. Springer, 2017.

\bibitem[Weyns(2019)]{weyns2019software}
Danny Weyns.
\newblock Software engineering of self-adaptive systems.
\newblock In \emph{Handbook of Software Engineering}, pages 399--443. Springer,
  2019.

\bibitem[Weyns et~al.(2012{\natexlab{a}})Weyns, Malek, and
  Andersson]{weyns2012forms}
Danny Weyns, Sam Malek, and Jesper Andersson.
\newblock {FORMS:} unifying reference model for formal specification of
  distributed self-adaptive systems.
\newblock \emph{ACM Transactions on Autonomous and Adaptive Systems (TAAS)},
  7\penalty0 (1):\penalty0 1--61, 2012{\natexlab{a}}.

\bibitem[Bruni et~al.(2012)Bruni, Corradini, Gadducci, Lafuente, and
  Vandin]{bruni2012conceptual}
Roberto Bruni, Andrea Corradini, Fabio Gadducci, Alberto~Lluch Lafuente, and
  Andrea Vandin.
\newblock A conceptual framework for adaptation.
\newblock In \emph{International Conference on Fundamental Approaches to
  Software Engineering}, pages 240--254. Springer, 2012.

\bibitem[Broy et~al.(2009)Broy, Leuxner, Sitou, Spanfelner, and
  Winter]{broy2009formalizing}
Manfred Broy, Christian Leuxner, Wassiou Sitou, Bernd Spanfelner, and Sebastian
  Winter.
\newblock Formalizing the notion of adaptive system behavior.
\newblock In \emph{Proceedings of the 2009 {ACM} Symposium on Applied Computing
  (SAC)}, pages 1029--1033. {ACM}, 2009.
\newblock \doi{10.1145/1529282.1529508}.

\bibitem[Lints(2012)]{lints2012essentials}
Taivo Lints.
\newblock The essentials of defining adaptation.
\newblock \emph{IEEE Aerospace and Electronic Systems Magazine}, 27\penalty0
  (1):\penalty0 37--41, 2012.

\bibitem[Weyns(2020)]{weyns2020introduction}
Danny Weyns.
\newblock \emph{An Introduction to Self-adaptive Systems: A Contemporary
  Software Engineering Perspective}.
\newblock John Wiley \& Sons, 2020.

\bibitem[Weyns(2017)]{weyns2017software}
Danny Weyns.
\newblock Software engineering of self-adaptive systems: an organised tour and
  future challenges.
\newblock \emph{Chapter in Handbook of Software Engineering}, 2017.

\bibitem[Weyns and Ahmad(2013)]{weyns2013claims}
Danny Weyns and Tanvir Ahmad.
\newblock Claims and evidence for architecture-based self-adaptation: a
  systematic literature review.
\newblock In \emph{European Conference on Software Architecture}, pages
  249--265. Springer, 2013.

\bibitem[Muccini et~al.(2016)Muccini, Sharaf, and Weyns]{muccini2016self}
Henry Muccini, Mohammad Sharaf, and Danny Weyns.
\newblock Self-adaptation for cyber-physical systems: a systematic literature
  review.
\newblock In \emph{Proceedings of the 11th international symposium on software
  engineering for adaptive and self-managing systems}, pages 75--81, 2016.

\bibitem[Mac{\'\i}as-Escriv{\'a} et~al.(2013)Mac{\'\i}as-Escriv{\'a}, Haber,
  Del~Toro, and Hernandez]{macias2013self}
Frank~D Mac{\'\i}as-Escriv{\'a}, Rodolfo Haber, Raul Del~Toro, and Vicente
  Hernandez.
\newblock Self-adaptive systems: A survey of current approaches, research
  challenges and applications.
\newblock \emph{Expert Systems with Applications}, 40\penalty0 (18):\penalty0
  7267--7279, 2013.

\bibitem[Krupitzer et~al.(2018)Krupitzer, Pfannem{\"u}ller, Voss, and
  Becker]{krupitzer2018comparison}
Christian Krupitzer, Martin Pfannem{\"u}ller, Vincent Voss, and Christian
  Becker.
\newblock Comparison of approaches for developing self-adaptive systems.
\newblock 2018.

\bibitem[Quin et~al.(2021)Quin, Weyns, and Gheibi]{quin2021decentralized}
Federico Quin, Danny Weyns, and Omid Gheibi.
\newblock Decentralized self-adaptive systems: {A} mapping study.
\newblock In \emph{16th International Symposium on Software Engineering for
  Adaptive and Self-Managing Systems, SEAMS@ICSE 2021}, pages 18--29. {IEEE},
  2021.
\newblock \doi{10.1109/SEAMS51251.2021.00014}.

\bibitem[Kitchenham and Charters(2007)]{kitchenham2007guidelines}
Barbara Kitchenham and Stuart Charters.
\newblock Guidelines for performing systematic literature reviews in software
  engineering.
\newblock Technical Report EBSE 2007-001, Keele University and Durham
  University Joint Report, 2007.
\newblock URL
  \url{http://www.dur.ac.uk/ebse/resources/Systematic-reviews-5-8.pdf}.

\bibitem[Kitchenham et~al.(2010)Kitchenham, Pretorius, Budgen, Brereton,
  Turner, Niazi, and Linkman]{kitchenham2010systematic}
Barbara Kitchenham, Rialette Pretorius, David Budgen, O~Pearl Brereton, Mark
  Turner, Mahmood Niazi, and Stephen Linkman.
\newblock Systematic literature reviews in software engineering--a tertiary
  study.
\newblock \emph{Information and software technology}, 52\penalty0 (8):\penalty0
  792--805, 2010.

\bibitem[Wohlin(2014)]{wohlin2014guidelines}
Claes Wohlin.
\newblock Guidelines for snowballing in systematic literature studies and a
  replication in software engineering.
\newblock In \emph{Proceedings of the 18th international conference on
  evaluation and assessment in software engineering}, pages 1--10, 2014.

\bibitem[Weyns et~al.(2012{\natexlab{b}})Weyns, Iftikhar, De~La~Iglesia, and
  Ahmad]{weyns2012survey}
Danny Weyns, M~Usman Iftikhar, Didac~Gil De~La~Iglesia, and Tanvir Ahmad.
\newblock A survey of formal methods in self-adaptive systems.
\newblock In \emph{Proceedings of the Fifth International C* Conference on
  Computer Science and Software Engineering}, pages 67--79, 2012{\natexlab{b}}.

\bibitem[{Haesevoets} et~al.(2009){Haesevoets}, {Weyns}, {Holvoet}, and
  {Joosen}]{Haesevoets2009}
R.~{Haesevoets}, D.~{Weyns}, T.~{Holvoet}, and W.~{Joosen}.
\newblock A formal model for self-adaptive and self-healing organizations.
\newblock In \emph{2009 ICSE Workshop on Software Engineering for Adaptive and
  Self-Managing Systems}, pages 116--125. {IEEE}, May 2009.
\newblock \doi{10.1109/SEAMS.2009.5069080}.

\bibitem[{Hachicha} et~al.(2016){Hachicha}, {Dammak}, {Halima}, and
  {Kacem}]{Hachicha2016}
M.~{Hachicha}, E.~{Dammak}, R.~B. {Halima}, and A.~H. {Kacem}.
\newblock A correct by construction approach for modeling and formalizing
  self-adaptive systems.
\newblock In \emph{2016 17th IEEE/ACIS International Conference on Software
  Engineering, Artificial Intelligence, Networking and Parallel/Distributed
  Computing (SNPD)}, pages 379--384, May 2016.
\newblock \doi{10.1109/SNPD.2016.7515928}.

\bibitem[{Hachicha} et~al.(2018){Hachicha}, {Halima}, and
  {Kacem}]{Hachicha2018}
M.~{Hachicha}, R.~B. {Halima}, and A.~H. {Kacem}.
\newblock Formalizing compound {MAPE} patterns for decentralized control in
  self-adaptive systems.
\newblock In \emph{2018 12th International Conference on Research Challenges in
  Information Science (RCIS)}, pages 1--10, May 2018.
\newblock \doi{10.1109/RCIS.2018.8406680}.

\bibitem[Laddaga(1997)]{laddaga1997self}
R~Laddaga.
\newblock Self-adaptive software {DARPA BAA} 98-12, 1997.

\bibitem[Lewis et~al.(2011)Lewis, Chandra, Parsons, Robinson, Glette, Bahsoon,
  Torresen, and Yao]{lewis2011survey}
Peter~R Lewis, Arjun Chandra, Shaun Parsons, Edward Robinson, Kyrre Glette,
  Rami Bahsoon, Jim Torresen, and Xin Yao.
\newblock A survey of self-awareness and its application in computing systems.
\newblock In \emph{2011 Fifth IEEE Conference on Self-Adaptive and
  Self-Organizing Systems Workshops}, pages 102--107. IEEE, 2011.

\bibitem[Kounev et~al.(2017)Kounev, Lewis, Bellman, et~al.]{kounev2017notion}
Samuel Kounev, Peter Lewis, Kirstie~L Bellman, et~al.
\newblock The notion of self-aware computing.
\newblock In \emph{Self-Aware Computing Systems}, pages 3--16. Springer, 2017.

\bibitem[Psaier and Dustdar(2011)]{psaier2011survey}
Harald Psaier and Schahram Dustdar.
\newblock A survey on self-healing systems: approaches and systems.
\newblock \emph{Computing}, 91\penalty0 (1):\penalty0 43--73, 2011.

\bibitem[Ghosh et~al.(2007)Ghosh, Sharman, Rao, and Upadhyaya]{ghosh2007self}
Debanjan Ghosh, Raj Sharman, H~Raghav Rao, and Shambhu Upadhyaya.
\newblock Self-healing systems--survey and synthesis.
\newblock \emph{Decision support systems}, 42\penalty0 (4):\penalty0
  2164--2185, 2007.

\bibitem[Villegas et~al.(2011)Villegas, M{\"u}ller, Tamura, Duchien, and
  Casallas]{villegas2011framework}
Norha~M Villegas, Hausi~A M{\"u}ller, Gabriel Tamura, Laurence Duchien, and
  Rubby Casallas.
\newblock A framework for evaluating quality-driven self-adaptive software
  systems.
\newblock In \emph{Proceedings of the 6th international symposium on Software
  engineering for adaptive and self-managing systems}, pages 80--89, 2011.

\bibitem[Weyns et~al.(2012{\natexlab{c}})Weyns, Iftikhar, Malek, and
  Andersson]{weyns2012claims}
Danny Weyns, M~Usman Iftikhar, Sam Malek, and Jesper Andersson.
\newblock Claims and supporting evidence for self-adaptive systems: A
  literature study.
\newblock In \emph{2012 7th International Symposium on Software Engineering for
  Adaptive and Self-Managing Systems (SEAMS)}, pages 89--98. IEEE,
  2012{\natexlab{c}}.

\bibitem[Zhang and Cheng(2006)]{Zhang2006Modelbaseddevelopment}
Ji~Zhang and Betty H.~C. Cheng.
\newblock Model-based development of dynamically adaptive software.
\newblock In \emph{Proceeding of the 28th international conference on Software
  engineering - {ICSE} {\textquotesingle}06}. {ACM} Press, 2006.
\newblock \doi{10.1145/1134285.1134337}.

\bibitem[Allen et~al.(1998)Allen, Douence, and Garlan]{allen1998specifying}
Robert Allen, Remi Douence, and David Garlan.
\newblock Specifying and analyzing dynamic software architectures.
\newblock In \emph{International Conference on Fundamental Approaches to
  Software Engineering}, pages 21--37. Springer, 1998.

\bibitem[Bradbury et~al.(2004)Bradbury, Cordy, Dingel, and
  Wermelinger]{bradbury2004survey}
Jeremy~S Bradbury, James~R Cordy, Juergen Dingel, and Michel Wermelinger.
\newblock A survey of self-management in dynamic software architecture
  specifications.
\newblock In \emph{Proceedings of the 1st ACM SIGSOFT workshop on Self-managed
  systems}, pages 28--33, 2004.

\bibitem[{Klarl}(2015)]{Klarl2015}
A.~{Klarl}.
\newblock Engineering self-adaptive systems with the role-based architecture of
  helena.
\newblock In \emph{2015 IEEE 24th International Conference on Enabling
  Technologies: Infrastructure for Collaborative Enterprises}, pages 3--8, June
  2015.
\newblock \doi{10.1109/WETICE.2015.32}.

\bibitem[Qureshi et~al.(2011)Qureshi, Jureta, and Perini]{Qureshi2011}
Nauman~A Qureshi, Ivan~J Jureta, and Anna Perini.
\newblock Requirements engineering for self-adaptive systems: Core ontology and
  problem statement.
\newblock In \emph{International Conference on Advanced Information Systems
  Engineering}, pages 33--47. Springer, 2011.

\bibitem[{Arcaini} et~al.(2015){Arcaini}, {Riccobene}, and
  {Scandurra}]{Arcaini2015a}
P.~{Arcaini}, E.~{Riccobene}, and P.~{Scandurra}.
\newblock Modeling and analyzing {MAPE-K} feedback loops for self-adaptation.
\newblock In \emph{2015 IEEE/ACM 10th International Symposium on Software
  Engineering for Adaptive and Self-Managing Systems}, pages 13--23. {IEEE},
  May 2015.
\newblock \doi{10.1109/SEAMS.2015.10}.

\bibitem[Iglesia and Weyns(2015)]{Iglesia2015}
Didac Gil De~La Iglesia and Danny Weyns.
\newblock {MAPE-K} formal templates to rigorously design behaviors for
  self-adaptive systems.
\newblock \emph{ACM Trans. Auton. Adapt. Syst.}, 10\penalty0 (3), September
  2015.
\newblock \doi{10.1145/2724719}.

\bibitem[Bucchiarone and Mongiello(2019)]{bucchiarone2019ten}
Antonio Bucchiarone and Marina Mongiello.
\newblock Ten years of self-adaptive systems: From dynamic ensembles to
  collective adaptive systems.
\newblock In \emph{From Software Engineering to Formal Methods and Tools, and
  Back}, pages 19--39. Springer, 2019.

\bibitem[Gheibi et~al.(2021)Gheibi, Weyns, and Quin]{gheibi2021applying}
Omid Gheibi, Danny Weyns, and Federico Quin.
\newblock Applying machine learning in self-adaptive systems: A systematic
  literature review.
\newblock \emph{ACM Transactions on Autonomous and Adaptive Systems (TAAS)},
  15\penalty0 (3):\penalty0 1--37, 2021.

\bibitem[Lemos et~al.(2013)Lemos, Giese, M{\"u}ller, et~al.]{lemos2013software}
Rog{\'e}rio~de Lemos, Holger Giese, Hausi~A M{\"u}ller, et~al.
\newblock Software engineering for self-adaptive systems: A second research
  roadmap.
\newblock In \emph{Software engineering for self-adaptive systems II}, pages
  1--32. Springer, 2013.

\bibitem[Salehie and Tahvildari(2009)]{salehie2009self}
Mazeiar Salehie and Ladan Tahvildari.
\newblock Self-adaptive software: Landscape and research challenges.
\newblock \emph{ACM transactions on autonomous and adaptive systems (TAAS)},
  4\penalty0 (2):\penalty0 1--42, 2009.

\bibitem[Elhabbash et~al.(2019)Elhabbash, Salama, Bahsoon, and
  Tino]{elhabbash2019self}
Abdessalam Elhabbash, Maria Salama, Rami Bahsoon, and Peter Tino.
\newblock Self-awareness in software engineering: A systematic literature
  review.
\newblock \emph{ACM Transactions on Autonomous and Adaptive Systems (TAAS)},
  14\penalty0 (2):\penalty0 1--42, 2019.

\bibitem[Kounev et~al.(2014)Kounev, Brosig, and Huber]{kounev2014descartes}
Samuel Kounev, Fabian Brosig, and Nikolaus Huber.
\newblock The descartes modeling language.
\newblock Technical report, Institut f{\"u}r Informatik, University of
  Wuerzburg, Germany, 2014.

\bibitem[Petrovska(2021)]{petrovska2021self}
Ana Petrovska.
\newblock Self-awareness as a prerequisite for self-adaptivity in computing
  systems.
\newblock In \emph{2021 IEEE International Conference on Autonomic Computing
  and Self-Organizing Systems Companion (ACSOS-C)}, pages 146--149. IEEE, 2021.

\end{thebibliography}
\end{document}